\documentclass[epj]{svjour}
\usepackage{graphics}
\begin{document}
  \title{\boldmath  Reaction   ${\bf{pp\to  pp  \pi\pi\pi   }}$  as  a
background for hadronic decays of the $\eta^{\prime}$ meson.  }
\author{
%First author\inst{1} \and Second author\inst{2}% etc
 M. J. Zieli{\'n}ski\inst{1} \inst{,2}
 P. Moskal\inst{1}\inst{,2} \and
 A. Kup{\'s}{\'c}\inst{3} \inst{,4}
}                     % Do not remove
\institute{
Institute of Physics, Jagellonian University, PL-30-059 Cracow, Poland
 \and
Nuclear Physics Institute, Research Center J{\"u}lich, D-52425 J\"{u}lich, Germany
 \and
Department of Physics and Astronomy, Uppsala University, Box 516, 75121 Uppsala, Sweden
 \and 
High Energy Physics Department, The Andrzej So\l{}tan Institute for Nuclear Studies,
Ho\.{z}a 69, PL-00681, Warsaw, Poland
}
\date{Received: \today / Revised version: date}

\abstract{Isospin violating hadronic decays of the $\eta$ and $\eta'$
  mesons into 3$\pi$ mesons are driven by a term in the QCD Lagrangian
  proportional to the mass difference  of the $d$ and $u$ quarks.  The
  source giving large  yield of the mesons for  such decay studies are
  $pp$    interactions   close    to   the    respective   kinematical
  thresholds. 
  The  most important physics background  for $\eta,\eta'\to\pi\pi\pi$ is coming
  from direct three pion production  reactions.  In case of the $\eta$
  meson  the background for  the decays  is relatively  low ($\approx$
  10\%).  The purpose of this article is to provide an estimate of the
  direct pion production background for the $\eta'\to 3\pi$ decays.  
  Using the inclusive data from  COSY-11 experiment we have  extracted
  differential cross section for the $pp\to pp$-multipion production 
reactions with the invariant
  mass of the pions equal to the $\eta'$ meson mass and 
  estimated an upper  limit for the  signal to background ratio for 
  studies of the $\eta'\to\pi^+\pi^-\pi^0$ decay. 
\PACS{
      {13.60.Le}{Meson production}   \and
      {13.75.-n}{Hadron-induced low- and intermediate-energy reactions and
             scattering (energy $\le$ 10 GeV)} \and
      {13.85.Lg}{Total cross sections}   \and
      {25.40.-h}{Nucleon-induced reactions}   \and
      {29.20.Dh}{Storage rings}
     } % end of PACS codes
} %end of abstract
\maketitle

\section{Motivation}
\label{Introduction} 

\subsection{\boldmath Three pion decays of the $\eta$ and $\eta'$ mesons}
The $\eta$  and $\eta'$  decays into three  pions violate  isospin and
occur only due to $u$ and  $d$ quark mass difference.  The decay width
is  sensitive   to  the  mass   difference:  $\Gamma_{\eta(\eta')  \to
\pi^+\pi^-\pi^0}\propto \Gamma_0\cdot  (m_d-m_u)^2$, where $\Gamma_0$
term can be  calculated in the  isospin limit $m_d=m_u$.  The  decays might
provide  a  precise  constraint   for  the  light  quark  mass  ratios~\cite{Leutwyler:1996qg}.

In the case of the  $\eta^{\prime}$ meson the existence of the isospin
conserving decays into three pseudoscalars ($\pi\pi\eta$) implies that
instead of the decay width for the isospin violating $\pi\pi\pi$ channel 
one can measure ratio:
\begin{equation}
\frac{BR( \eta'\to \pi\pi\pi)}{BR( \eta'\to \pi\pi\eta)},
\label{eqn:GTR}
\end{equation}
as it was proposed  by Gross, Treiman and Wilczek~\cite{Gross:1979ur}.
Such  measurement  is  self   contained  since  it  does  not  require
normalization for the  partial decay width  from
other  experiments.   Additionally, simultaneous  measurement of
the two  decays modes,  with similar final  states, ensures  that many
systematic uncertainties  will cancel.  
The $\eta'\to\pi\pi\pi$ decays provides also a very sensitive test 
of the Chiral Perturbation Theory (CHPT) framework~\cite{Weinberg:1978kz,Gasser:1983yg} 
extensions to the $\eta'$ meson. Due  to the  large mass  of  the $\eta'$
meson, the decays  are strongly influenced by light  vector and scalar
meson  resonances.   Those  decays  cannot be  studied  using standard  CHPT
methods.   An elegant  method for  accounting for  the nonperturbative
effects   in  two   pseudoscalar  meson   interactions  is   given  by
unitarization procedure  of the one  loop CHPT result. For  example in
the theoretical  studies of  the $\eta$ and  $\eta'$ meson  decays the
rescattering of  any pair of  the pseudoscalar mesons is  described by
Bethe-Salpether                                                equations
\cite{Oller:1998zr,Oller:1998hw,Beisert:2003zd}.   The  parameters  of
the interactions  are obtained by fits to  the pseudoscalar scattering
data. Predictions for many $\eta$ and $\eta'$ decays were given using the 
above technique within so called {\it chiral   unitary  approach}~\cite{Borasoy:2006uv}.

Experimentally determined value of the branching  ratio of the $\eta^{\prime}\to 3\pi^0$ decay
is (1.56$\pm$0.26)$\times$10$^{-3}$ \cite{Binon:1984fe}.
The decay  into $\pi^+\pi^-\pi^0$ was observed in 2009 for the first
time by the CLEO collaboration  and the branching ratio was determined
to   $(37^{+11}_{-9}\pm  4)\times   10^{-4}$~\cite{Naik:2008tb}.   The
result is  in strong disagreement  with the value  10$^{-2}$ predicted
within  the chiral   unitary  approach.   Much  more
improvement on the theory and  experiment side is needed to understand
the three pion decays of the $\eta'$ meson.

\subsection{\boldmath Experiments using $pp\to pp\eta',\eta$ reaction}

\begin{figure}[t!]
\vspace{-1.5cm}
\resizebox{0.47\textwidth}{!}{%
 \includegraphics{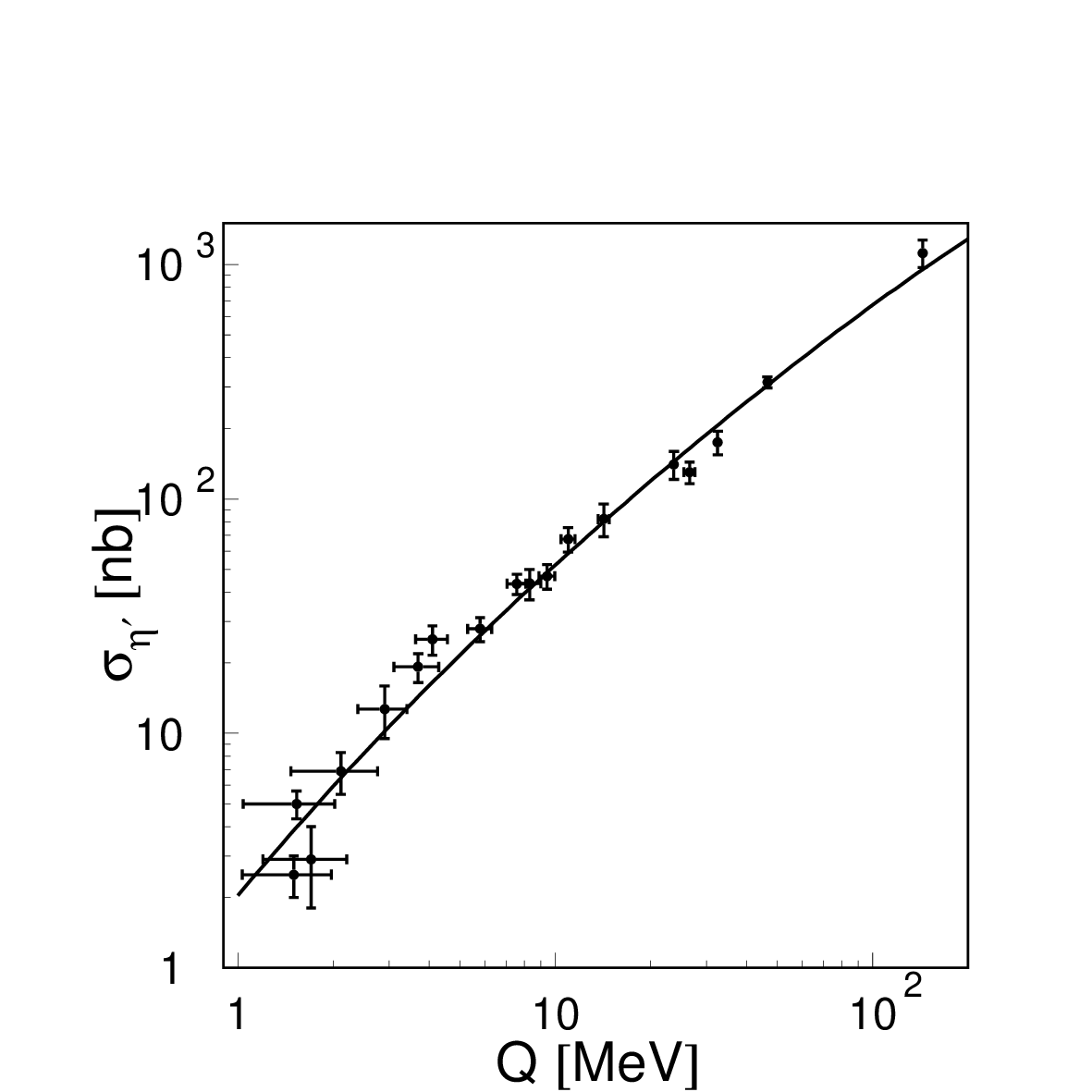}}
\caption{   Compilation of the   data for   the   $pp\to
  pp\eta^{\prime}$ reaction cross section  from COSY-11, DISTO and
  SPESIII
  measurements~\cite{Moskal:1998pc,Moskal:2000gj,Khoukaz:2004si,Hibou:1998de,Balestra:2000ic}.
  The  solid  line is  parameterization of  the data
  using formula (\ref{przekrojparam}).  }
\label{przekrojeta}
\end{figure}

The   mesons  for   the  decay   studies  are   produced   in  $\gamma
p$~\cite{Unverzagt:2008ny,Unverzagt:2009vm},
$pp$~\cite{Moskal:2003gt,Moskal:2002jm},
$\pi^-p$~\cite{Starostin:2007zz,Arndt:2006hj},
$pd$~\cite{Smyrski:2007nu,Mersmann:2007gw,PiskorIgnatowicz:2006nd}
or $e^+e^-$~\cite{AmelinoCamelia:2010me,Li:2009jd} interactions.  For the
studies  at  light ion  storage  rings  as  COSY the  $pp\to  pp\eta'$
reaction close  to threshold  seems to be  most promising.   The cross
section  for  $pp\to pp\eta^{\prime}$  reaction  was  measured by  the
COSY-11~\cite{Moskal:1998pc,Moskal:2000gj,Khoukaz:2004si,Klaja:2010vy},
SPESIII~\cite{Hibou:1998de}      and      DISTO~\cite{Balestra:2000ic}
collaborations.  In  Fig.~\ref{przekrojeta} the experimental  data are
compared to  the analytical parameterization derived  by F\"{a}ldt and
Wilkin~\cite{Faeldt:1996na,Faldt:1997jm}  which   takes  into  account
final state interaction of the protons:
\begin{equation}
\sigma^{tot}_{\eta^{\prime}}(Q) = 
C \frac{Q^2}{m_p p_{LAB}}\frac{1}{\left(1 + \sqrt{ 1 + \frac{Q}{\epsilon}}\right)^2},
\label{przekrojparam}
\end{equation}
where $Q$ denotes the excess energy, $p_{LAB}$ beam momentum and $m_p$
proton mass.   
In comparison to the $pp$ interaction the $p-\eta'$ interaction is negligible~\cite{Moskal:2000pu}.
The $C$ and  $\epsilon$ free parameters were determined
by a fit  to the experimental  data \cite{Moskal:2006qq}: $\epsilon=0.62
\pm 0.13$~MeV and $C=42\pm7$~mb.

\begin{figure}[t!]
\hspace{0.55cm}
\resizebox{0.40\textwidth}{!}{%
 \includegraphics{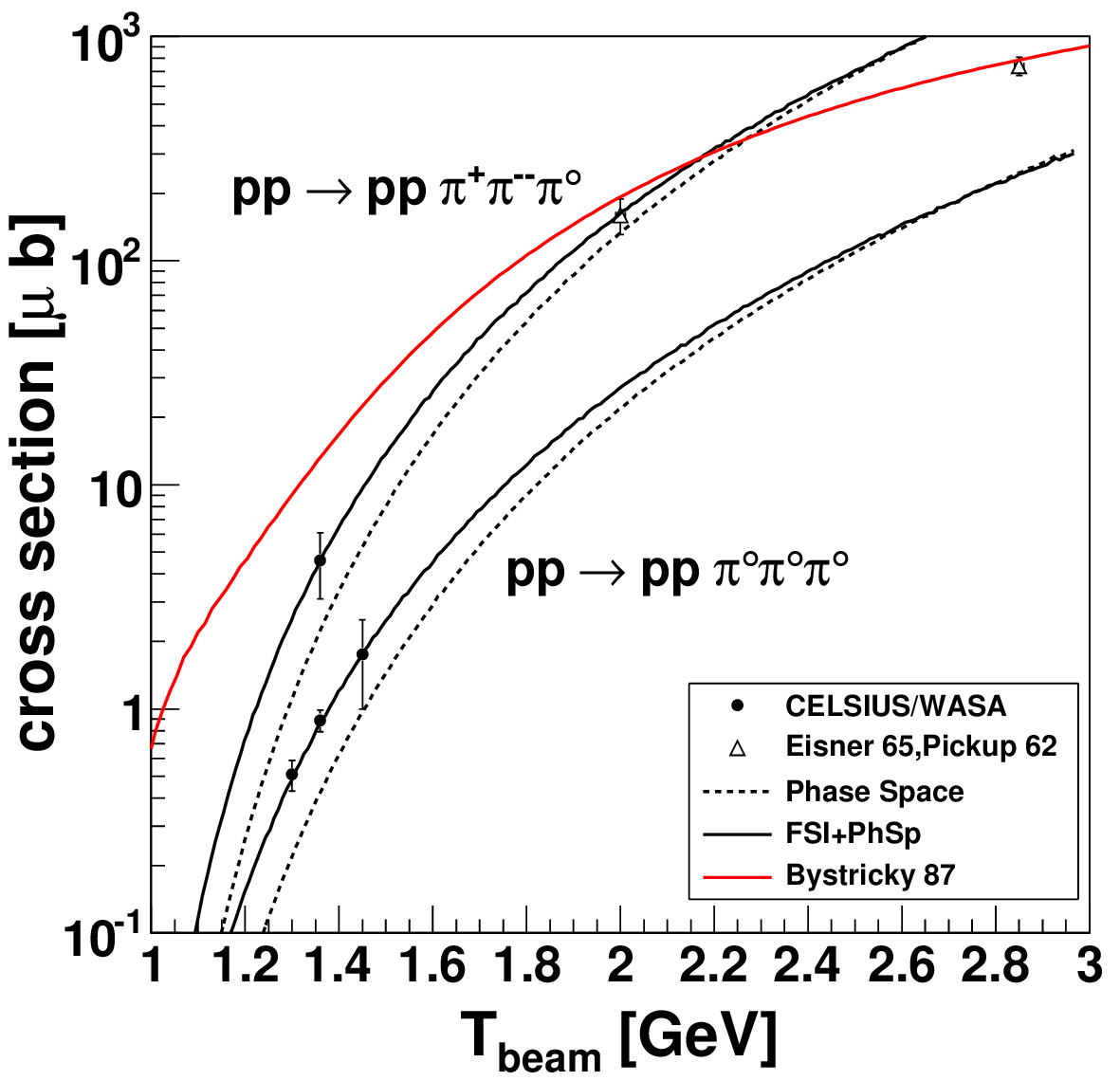}}
\caption{\label{pp3pi}
Total cross  section for three  pion production:
data       and      parameterizations.       The data       are      from
\cite{Hart:1962aa,Pickup:1962aa,Pickup:1962ab,Eisner:1965aa,Pauly:2006pm}.
The parameterizations  are from  Bystricky et  al.  \cite{Bystricky:1987yq} and 
F\"aldt and  Wilkin~\cite{Faeldt:1996na,Faldt:1997jm}.  The kinetic beam 
energy at thresholds for the
$\eta$   and  $\eta'$  production  is equal  to 1.255   GeV  and   2.404  GeV
respectively.}
\label{3picrossBystr}
\end{figure}

Experience from studies  of the  $\eta\to\pi\pi\pi$ decays with $\eta$ produced in  $pp\to pp\eta$
reaction  at beam energies  1.30--1.45 GeV  carried out  by the CELSIUS/WASA
collaboration shows that background  from direct three pion production
is  about 15\%  for the  $\pi^+\pi^-\pi^0$  and about 5\%  for the  $3\pi^0$
channel~\cite{Pauly:2006pm,Bashkanov:2007iy}.  This allows for precise
study  of the  $\eta$ decays  providing a  large number  of  events is
collected. The  production cross section  for the $\eta$ mesons  in $pp$
colisions~\cite{Bergdolt:1993xc,Chiavassa:1994ru,Calen:1996mn,Smyrski:1999jc,Moskal:2003gt}
is about 30 times larger  than the cross section for the
$\eta^{\prime}$
meson~\cite{Moskal:1998pc,Moskal:2000gj,Khoukaz:2004si,Klaja:2010vy,Hibou:1998de,Balestra:2000ic}
at  the corresponding  excess energies.   At the  same time  the total
cross section for the direct three pion production increases about two
orders  of  magnitude between  $\eta$  and $\eta^{\prime}$  production
thresholds (Fig.~\ref{pp3pi}). For the  $pp\to pp\pi^+\pi^-\pi^0$ reaction total  cross section there
are only three experimental points in the beam kinetic energy range up
to 3 GeV.  The
cross section for the $pp\to pp\pi^0\pi^0\pi^0$ reaction near  the 
$\eta^{\prime}$ meson production threshold was
not  measured at all.   However,  based on the  statistical
model~\cite{Fermi:1950jd,Cerulus:1958gp,Cerulus:1960aa}   and  the  isobar
model  \cite{Lindenbaum:1957aa,Sternheimer:1961aa}  one expects  the
the  $pp\to pp\pi^0\pi^0\pi^0$ cross section  to be
$6-10$  times  lower  than for the  $pp\to  pp\pi^+\pi^-\pi^0$
reaction.   This  is  in   agreement  with  an  extrapolation  of  the
CELSIUS/WASA measurement  of the both reactions  close to $\eta$  meson production threshold.
For the  estimate of the background  for the three pion  decays of the
$\eta'$  mesons  instead  of  the  total cross  section  the  relevant
quantity  is the  value  of  the differential  cross  section for  the
invariant masses of the pions in  the range of the $\eta'$ meson mass.
This  quantity was  not  measured and  in  the present  paper we  will
provide an  estimate for the upper  limit of the  background using the
COSY-11 data where only outgoing protons were registered.

\section{General considerations}

Let us  consider an example analysis  chain leading to  a selection of
the  $\eta'\to\pi^{+}\pi^{-}\pi^{0}$ decay  from  the $pp\to  pp\eta'$
reaction.   In  the  first  step  all  tracks  are  reconstructed  and
particles  identified.  The  events  containing two  protons from  the
production process,  two charged pions  and two photons  are selected.
Now one can apply  energy-momentum conservation and select only events
consistent     with     $pp\to     pp\eta'\to     pp\pi^+\pi^-\pi^0\to
pp\pi^+\pi^-\gamma\gamma$ reaction hypothesis.   This procedure can be
most generally implemented  by kinematic fitting but one  can use also
some other method.  In the end the selection could be represented by a
region  in some  control variable  $\mu$. For  example $\mu$  could be
missing mass squared  or $\chi^2$ value of the  kinematic fit.  Within
the selected region,  in addition to the signal,  a contamination from
the background  events originating  from reactions which  have similar
final states  is unavoidable.   The most important  physics background
channels for  the discussed  case is the  direct $pp\to  pp\pi^+ \pi^-
\pi^0$  reaction  and  other   $\eta'$  decays  like  $\eta'\to  \pi^+
\pi^-\eta   \to   \pi^+   \pi^-\gamma\gamma$.   Candidates   for   the
identification variable (in addition to $\chi^2$ of the kinematic fit)
are in these  cases respectively: the missing mass  of the two protons
and the invariant mass of the two photons. In this article we focus on
the  first case:  estimate  the  signal to  background  ratio and  its
implications  for the  statistical  uncertainty of  the extraction  of
$BR(\eta'\to \pi^+ \pi^-\pi^0)$ value.

The signal to background ratio,  $N_S/N_B$, can be written as:
\begin{equation}
  \frac{N_S}{N_B}=\frac{\sigma_{\eta'}\cdot BR\cdot\varepsilon_S\cdot{\cal L}}
{\Delta\mu\cdot\rho_B\cdot\varepsilon_B\cdot{\cal L}},
\label{eq3}
\end{equation}
where the factors are: 
\begin{enumerate} 
\item $\sigma_{\eta'}$ -- the total cross section for the production reaction (here for $pp \to pp \eta^{\prime}$),
\item   $\rho_B$ -- the  differential   cross  section   for  the   direct  $\pi^{+}\pi^{-}\pi^{0}$ 
production  with the pions invariant  mass 
equal to the mass of the $\eta^{\prime}$ meson:
\begin{equation}
\rho_{B}    \equiv
\left.\frac{d\sigma_B}
{d\mu}\right|_{\mu=m_{\eta'}},
\end{equation}
\item $\varepsilon_S$, $\varepsilon_B$ -- acceptances and reconstruction efficiencies for the signal and the 
background,
\item $\Delta\mu$ -- a range of the missing mass used for the extraction of the signal it depends on the detector resolution,
\item $BR$ -- the measured 
branching ratio of the  $\eta^{\prime} \to \pi^{+}\pi^{-}\pi^{0}$ decay,
\item ${\cal L}$ -- stands for the integrated luminosity.
\end{enumerate}
The $N_S/N_B$ ratio depends on the beam energy through 
 $\sigma_{\eta'}$,  $\rho_B$,  the   missing  mass  resolution   and  the
detection  efficiencies.   Hereafter we  derive the
energy dependence of these quantities.

\section{Background estimate}

\begin{figure}[t!]
\vspace{-0.5cm}
\resizebox{0.47\textwidth\vspace{1.cm}}{!}{%
 \includegraphics{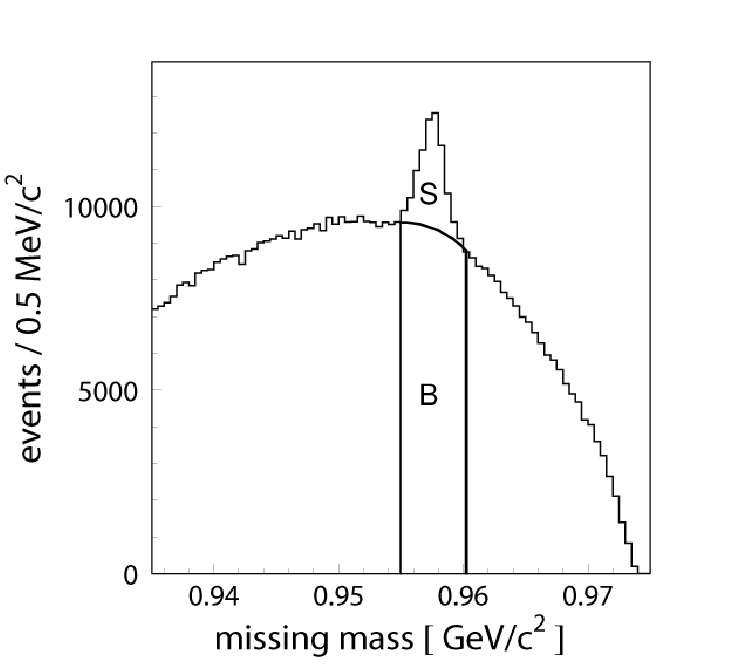}}
\vspace{-1.cm}
 \caption{ An example of the  missing mass distribution for the
   $pp\to ppX$  reaction from the  COSY-11 measurements at
   $Q = 15.5$ MeV~\cite{Moskal:2006qq}.  $S$ denotes the signal
   originating from  the $\eta^{\prime}$ meson production  and $B$ indicates
   the  background  under the  peak.   The  continuum originates  from
   direct production of two,  three and  more mesons.   It provides
   a conservative upper limit for  the background
   from the $pp\to pp\pi^{+}\pi^{-}\pi^{0}$ reaction.  }
\label{wykres_masy}
\end{figure}

The value of the $\rho_B$ cross section should be determined from
the $\pi^{+}\pi^{-}\pi^{0}$ invariant mass  distributions of the direct 
pion production reaction. However, there is  no data at beam energies  
near   the   $\eta^{\prime}$  meson threshold.  Therefore,  
we  estimate an upper limit  for the $\rho_B$
by  re-evaluating the  available
missing mass  spectra of  the $pp\to ppX$  reaction determined  by the
COSY-11 collaboration at several beam energies near the
the                        $\eta^{\prime}$  threshold~\cite{Moskal:1998pc,Moskal:2000gj,Khoukaz:2004si,Klaja:2010vy}.
In  Fig.~\ref{wykres_masy}  an  example of  COSY-11
reconstructed  missing  mass  distribution  at the $pp\to pp\eta'$ 
excess  energy, $Q$,  of
15.5~MeV is shown~\cite{Moskal:2006qq}.

Let us consider a measurement of  the two protons from the $pp\to ppX$
reaction,  where  there  are  no  constrains on  other  outgoing
particles.   In a first  approximation the  ratio of  the number  of the
background events  in a  slice under the  $\eta'$ peak in  the missing
mass spectrum to  the number of events in the peak  does not depend on
the detector acceptance for  the protons.  This assumption is valid
for example  if the $\eta^{\prime}$ meson and  the multipion reactions
are simulated according  to phase space or even  if an universal final
state          interaction         between          protons         is
introduced~\cite{Moskal:2003fi,Moskal:2005uh}.    The  assumption  may
break for  example if there will  be a significant  difference in four
momentum  transfer  distributions   between  $\eta^{\prime}$  and
multimeson  production.   However,  the  dependence of  the  production
amplitude on the  momentum transfer is very weak  near the kinematical
threshold.

\begin{figure}[t!]
\resizebox{0.47\textwidth}{!}{%
 \includegraphics{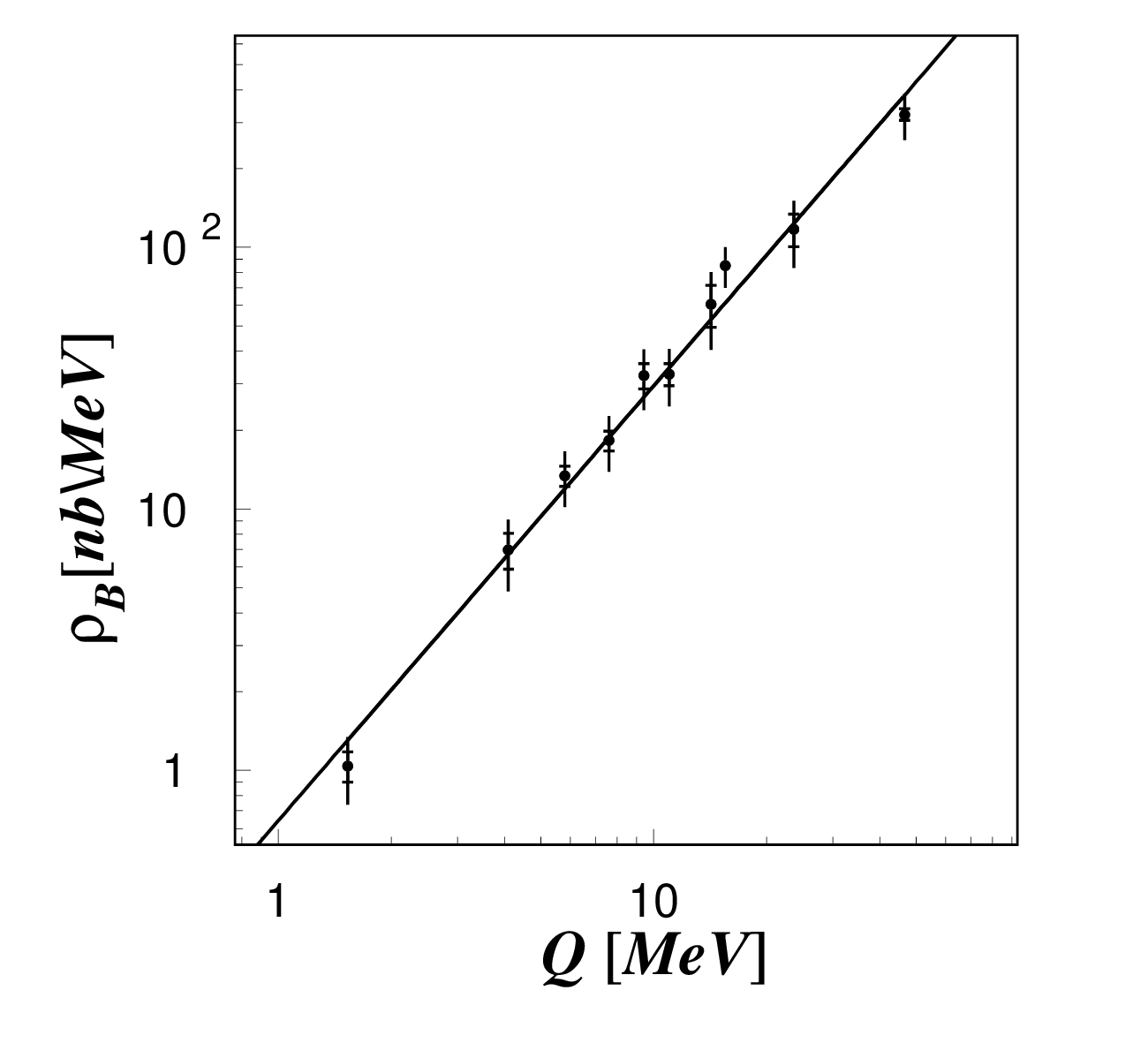}}
 \caption{Inclusive   differential  cross  section   for  multimeson
production derived from the COSY-11 data~\cite{Moskal:1998pc,Moskal:2000gj,Khoukaz:2004si,Klaja:2010vy}.  
Statistical and  systematic errors
are  seperated by the horizontal  bars.  The  superimposed line  shows the
function of Eqn.~\ref{eqn:rhcosy} fitted  to the  COSY-11 data.
\label{diffcross}
}
\end{figure}

The   differential  cross   section  $\rho_{B}$   for   the  background
originating from  all multimeson channels  was  determined from  the
COSY-11 data according to the formula:
\begin{equation}
\rho_B(Q) = \frac{N_{B}(Q)}{N_{S}(Q)} 
\frac{\sigma_{\eta^{\prime}}(Q)}{\Delta \mu } ,
\label{diff_cross_section}
\end{equation}
which is derived from the following expressions for the $N_{S}(Q)$ and
$N_B(Q)$:
\begin{eqnarray}
     N_{S}(Q) &    =&      \sigma_{\eta^{\prime}}(Q)     \cdot
\varepsilon(m_{\eta^{\prime}},Q)\cdot {\cal L},
\label{NS}\\
N_{B}(Q) &  = &  \rho_{B}(Q)    \cdot   \Delta   \mu   \cdot   \varepsilon
(m_{\eta^{\prime}},Q) \cdot {\cal L},
\label{NB}
\end{eqnarray}
where  N$_S$ stands  for  the number  of  the observed  events in  the
$\eta^{\prime}$  peak, N$_B$ number  of the  background events  in the
$\Delta    \mu$     slice    under    the     $\eta^{\prime}$    signal,
$\sigma_{\eta^{\prime}}$  denotes  the  total  cross section  for  the
$pp\to  pp\eta^{\prime}$  reaction described according to analitical formula
from Eq.\ref{przekrojparam},  $\epsilon$
denotes combined  acceptance and  detection efficiency of  the COSY-11
detector, which,  in a  very good approximation,  depends only  on the
mass    of    the    produced     system    and    on    the    excess
energy~\cite{Moskal:2003fi,Moskal:2005uh}, $\mathcal  L$ indicates the
integrated luminosity.  The width of the $\Delta \mu$ slice was selected
to contain nearly 100\% events of the signal peak.  
The  derived  values   of  $\rho_{B}$  as  a  function   of  $Q$  with
corresponding statistical and systematic errors are given in Table~\ref{tabela1}
and are shown in Fig.~\ref{diffcross}.
%\begin{center}
\begin{table}
\begin{center}
%\hspace{4.0cm}
\begin{tabular}{c|ccc}\hline\hline
Q             & $\rho_{B}$  & $\Delta \rho_{B}(stat.)$ & $\Delta \rho_{B}(syst.)$   \\ 
\mbox{} [MeV] &[nb/MeV]&[nb/MeV]& [nb/MeV]             \\ \hline\hline
1.53  & 1.04   & 0.14   & 0.16 \\
4.10  & 7.0   & 1.1   & 1.1 \\
5.80  & 13.4  & 1.2   & 2.0 \\
7.60  & 18.2  & 1.6   & 2.8 \\
9.42  & 32.3  & 3.6   & 4.9 \\
10.98 & 32.7  & 3.2   & 4.9 \\
14.21 & 60  & 11  & 9 \\
15.50 & 85  & 2.4   & 13\\
23.64 & 117 & 17  & 17\\
46.60 & 322  & 16  & 48\\ \hline\hline
\end{tabular}
\end{center}
\caption{Differential   cross   section   $\rho_{B}$  for   multimeson
production in proton-proton collisions as a function of the
$pp\to pp\eta'$ reaction
excess energy  $Q$.   The  $\rho_{B}$
values       were      extracted      from       the      experimental
data~\cite{Moskal:1998pc,Moskal:2000gj,Khoukaz:2004si,Klaja:2010vy}    using
Eq.~\ref{diff_cross_section}.}
\label{tabela1}
\end{table}
The systematic uncertainties are discussed in \cite{Moskal:1998pc,Moskal:2000gj,Khoukaz:2004si,Klaja:2010vy}.

The  $\rho_{B}$  dependence on the  excitation energy  is well
described by the following parameterization:
\begin{equation}
\rho_B=\alpha  (Q/Q_0)^{\beta}\label{eqn:rhcosy}
\end{equation}
where $Q_0=1$ MeV is the normalization factor and $\alpha$ and $\beta$ are the parameters.  The fit gives
$\alpha= 0.64 \pm 0.14$ nb/MeV and $\beta = 1.662 \pm 0.081$.  

\section{\boldmath Statistical uncertainty of the BR measurement}

Assuming that the shape of the  background is known and is not
correlated with the signal the relative statistical uncertainty of the
branching ratio can be expressed as:
\begin{equation}
\frac{\sigma(BR)}{BR}=\frac{\sigma(N_S)}{N_S}
= \frac{\sqrt{N_{S} + N_{B}} }{N_{S}}.
\label{blad_wzgledny_br}
\end{equation}
In our case  $N_S\ll N_B $ and we have:
\begin{equation}
\frac{\sigma(BR)}{BR}=\frac{\sqrt{N_B}}{N_S}
\label{eq10}
\end{equation} 
Taking into account Eq.~\ref{eq3} and Eq.~\ref{eq10} 
one gets the formula  for  the  statistical  uncertainty:
\begin{equation}
\sigma(BR) \le   \frac{\sqrt{\rho_B\cdot\Delta\mu\cdot\varepsilon_B}}
{\sigma_{\eta'}\cdot\varepsilon_S}\frac{1}{\sqrt{\mathcal L}}.
\end{equation}
From the derived expression  one sees that improved tagging resolution
($\Delta\mu$ decreased) helps only  if the detection efficiency is not
worsened.  

The  integrated luminosity  $\mathcal L$  can be  determined  from the
simultaneously            measured           decay           $\eta'\to
\pi^+\pi^-\eta\to\pi^+\pi^-\gamma\gamma$    with    well   established
branching  ratio.   Therefore   the  statistical  uncertainty  of  the
luminosity determination  can be  neglected and many  contributions to
the systematic  uncertainty in the  ratio of signal to  the monitoring
events will cancel.

\section{Feasibility for a large acceptance detector} 

As  an example  application of  the extracted  $\rho_B$ value  and the
formulas  derived  in  the   previous  sections  we  will  consider  a
determination   of    the   branching   ratio    for   the   $\eta'\to
\pi^+\pi^-\pi^0$ decay  using a large acceptance  detector.  The large
acceptance is  necessary for efficient identification  of all outgoing
particles.   After selection  of  the $pp\pi^{+}\pi^{-}\pi^{0}$  final
state  the  direct  tree  pion production  and  the  $\eta^{\prime}\to
\pi^{+}\pi^{-}\pi^{0}$  decay  can  be  distinguished  with  the  best
precision using the missing mass of the two forward emitted protons.

For the calculations of the missing mass resolution a typical beam and
target    parameters    available    at   the    Cooler    Synchrotron
COSY~\cite{Maier:1997zj}  were assumed:  beam momentum  spread $\Delta
p/p\approx  10^{-3}$ (FWHM),  perpendicular beam  profiles: horizontal
$\sigma_X$=2~mm, vertical  $\sigma_Y$=5~mm \cite{Moskal:2000em}.  As a
target a  hydrogen stream in  a cylinder with  diameter of 2.5  mm was
used.  Effective  energy resolution  of the forward  scattered protons
from   the   reaction  $pp\to   pp\eta'$   registered  using   plastic
scintillators is typically in the order of few percent (3\%).

\begin{figure}[h!]
\resizebox{0.47\textwidth}{!}{
\includegraphics{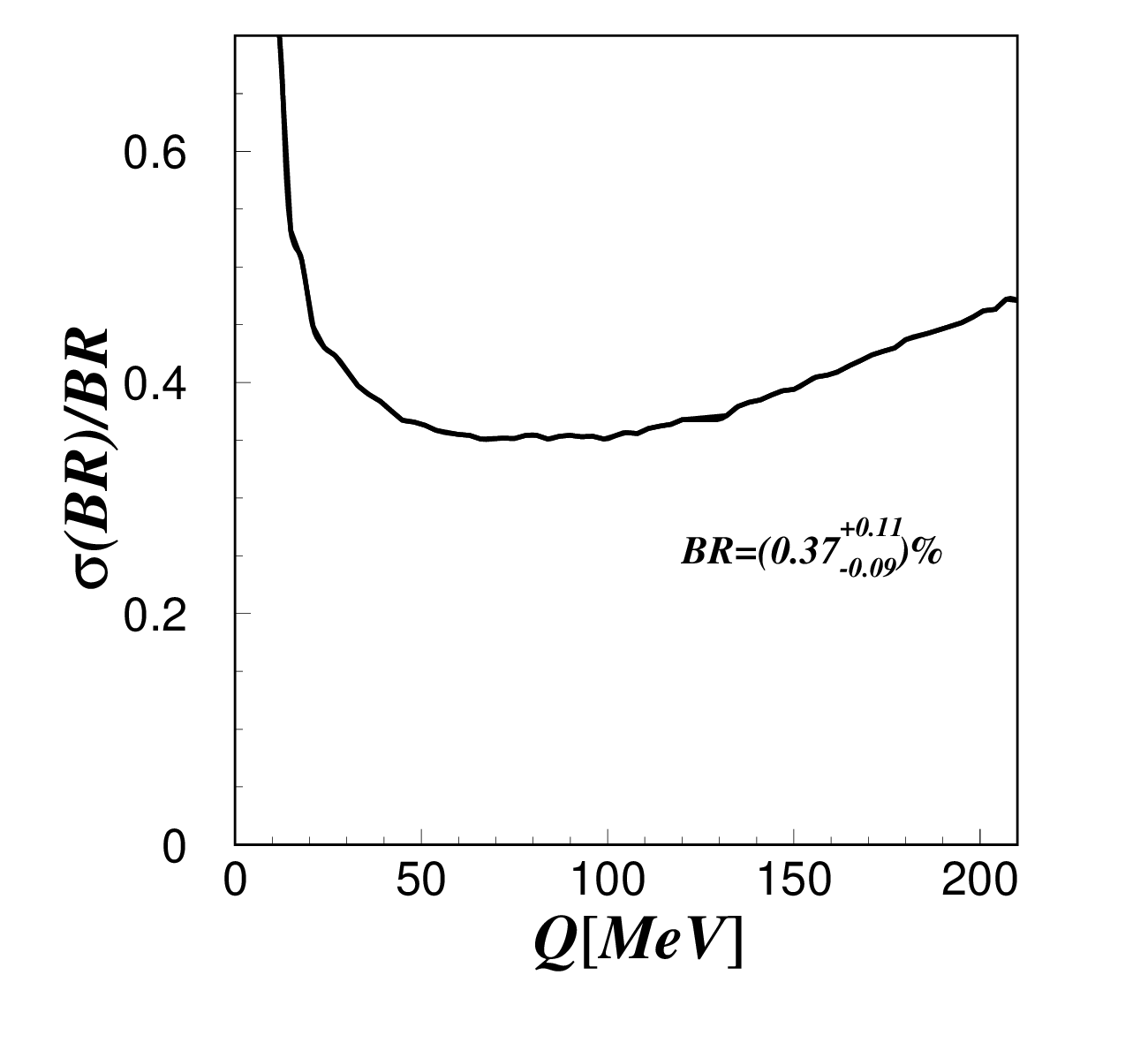}}
 \caption{
\label{fig:5} The relative accuracy of the upper limit of the  
$BR(\eta^{\prime}\to \pi^+\pi^-\pi^0$) as a
function  of the  excess energy  $Q$ for  the  $pp\to pp\eta^{\prime}$
reaction.}
\end{figure}
\begin{figure}[t!]
\vspace{-0.4cm}
\resizebox{0.45\textwidth}{!}{\includegraphics{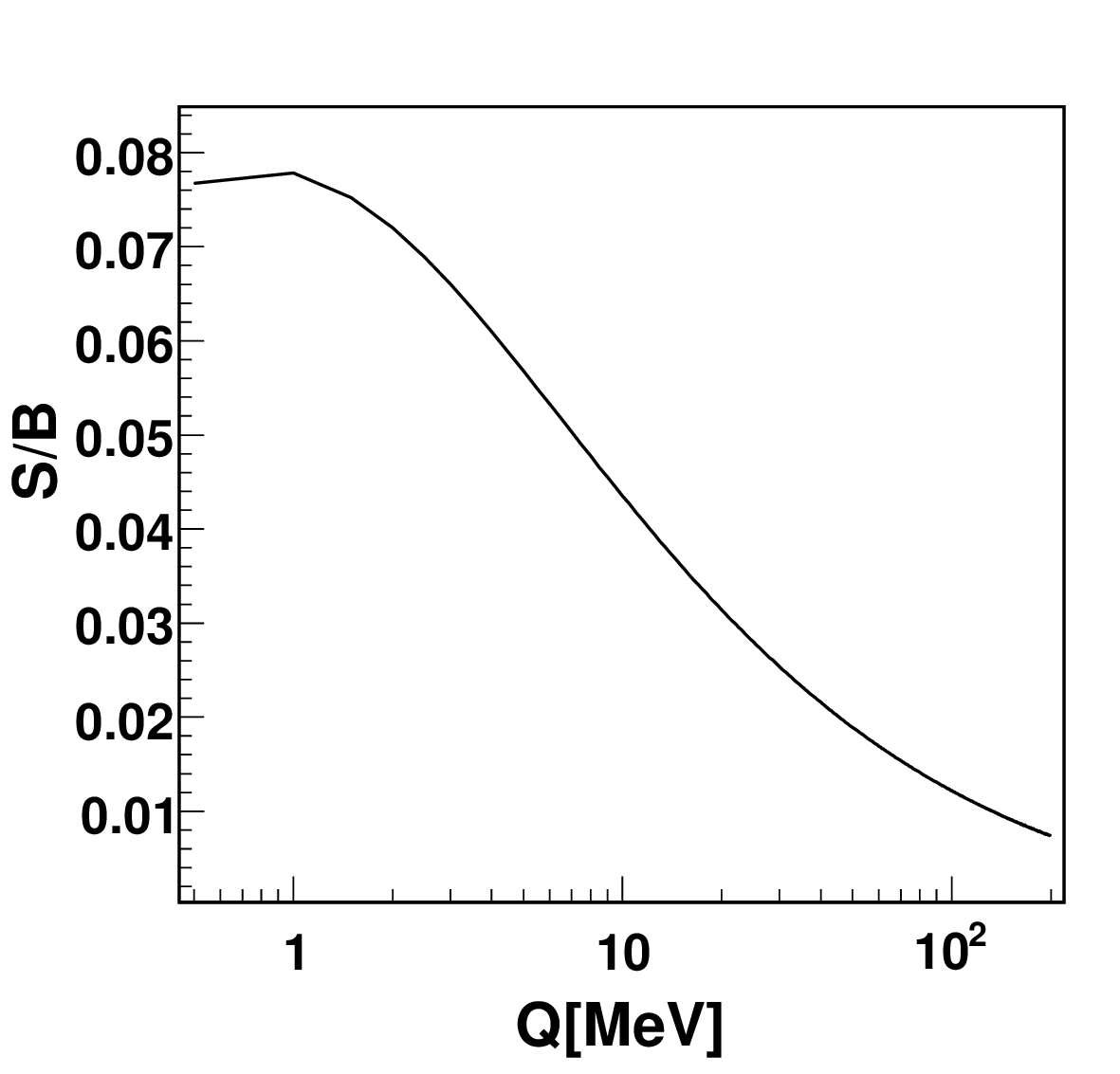}}
\vspace{0.3cm}
\caption{\label{fig:8}  The  signal to background ratio S:B calculated taking into 
account the natural width of the $\eta'$ meson.
}
\end{figure}

The determined $Q$  dependence of ${\sigma(BR)}/{BR}$ is shown
in Fig.~\ref{fig:5}, assuming     one     week     experiment    with
 luminosity     of
L~=~10$^{32}$cm$^{-2}$s$^{-1}$      and        value      of
$BR(\eta^{\prime}\to\pi^{+}\pi^{-}\pi^{0})$ = 0.37\%~\cite{Naik:2008tb}.
The  optimum  is
reached for the excess  energies between 50 and 100~MeV.
This  is a  general conclusion  for  a large
acceptance detection systems.
\begin{figure}[t!]
\resizebox{0.47\textwidth}{!}{
 \includegraphics{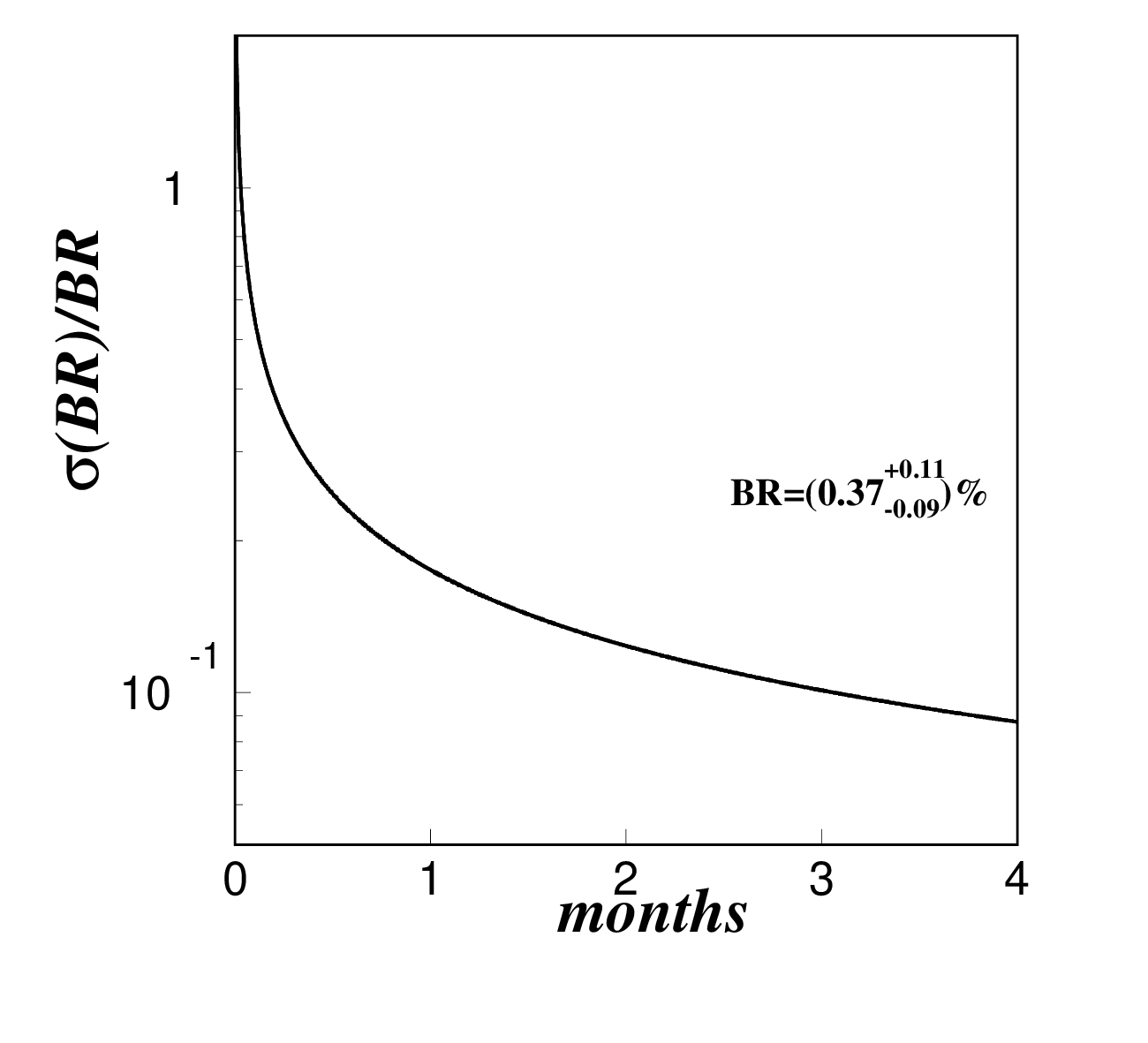}}
\caption{\label{fig:6}  The  relative  accuracy of  the upper limit of the 
branching ratio for the  decay $\eta^{\prime}\to \pi^+\pi^-\pi^0$ as a
function of measurement time in months.  
}
\end{figure}
The statistical uncertainty of  the branching ratio improves with time
of  the measurement  as  $1/\sqrt{t}$.  The  dependence  for the  beam
momentum of p$_{beam}$~=~3.45 GeV/c corresponding to the excess energy
Q~=~75 MeV is shown in  Fig.~\ref{fig:6}.  The plot indicates that for
the BR  equal 0.37\%  to achieve the  relative accuracy of  10\% would
require at  least two month  experiment.  However signal  to backround
ratio at  this energy will be  only about $10^{-3}$  what puts extreme
requirements  for   the  understanding  of   the  systematic  effects.
Therefore the  other strategy  for the experiment  would be to  find a
compromise  between the  statistical and  systematic  uncertanities by
going  to lower  excitation energies  where the  signal  to background
ratio increases (Fig.\ref{fig:8}).

An  additional source of  background, not  discussed here,  comes from
other    decays    of    $\eta'$    involving    similar    particles:
$\eta'\to\pi^+\pi^-\eta$ and  $\eta'\to\omega\gamma$.  This background
cannot be suppressed  using the missing mass method  and the invariant
masses of the decay products should be used instead.

\section{Summary} 

Using the COSY-11  data for the $pp\to ppX$  reaction near the $\eta'$
meson  production  threshold  we  extracted  an upper  limit  for  the
background   from   the   $pp\to   pp\pi^+\pi^-\pi^0$   reaction   for
$\eta'\to\pi^+\pi^-\pi^0$ decay studies.  This and the parameterization
of the total  cross section energy dependence for  the $pp\to pp\eta'$
reaction permit us  to estimate that two months of  the beam time with
average luminosity  of $10^{32}$cm$^{-2}$s$^{-1}$ would  be sufficient
to  reach statistical  accuracy of  the previous  experiments  for the
studies   of  $\eta'\to\pi^+\pi^-\pi^0$   using  a   large  acceptance
detector. However  since signal to  background ratio for  the optimal
energy is only  about $10^{-3}$ one may expect  much larger systematic
uncertainty. Since  even taking
the  natural width  of  the $\eta'$  meson  the $S:B$  ratio is  about
$10^{-2}$, this points to  inherent limitations of the $pp\to pp\eta'$
reaction   for   the  $\eta'\to\pi^+\pi^-\pi^0   $   decay  shown   in
Fig.~\ref{fig:8}.  Situation is  expected to be at least  one order of
magnitude better  for the $\eta'\to\pi^0\pi^0\pi^0$  case.  Also ''not
rare''  decays  $\eta'\to\pi\pi\eta$  can  be studied  in  the  $pp\to
pp\eta'$ reaction.

\begin{acknowledgement}
\begin{center}
{\bf Acknowledgments}
\end{center}

The work was partially supported
by the European Commission through the Research Infrastructures action of the Capacities Program. 
Call: FP7-INFRASTRUCTURES-2008-1, Grant Agreement
No. 227431, by the PrimeNet, by the FFE
grants from the Research Center J\"{u}lich, 
by the MPD programme of Foundation for Polish Science through structural funds of the European Union and 
by DOCTUS programme of Ma\l{}opolskie Centre of Entrepreneurship through structural funds of the European Union.
\end{acknowledgement}

\bibliographystyle{JHEP-2}
\bibliography{pi3}
\end{document}